\documentclass{pasj00}
\usepackage{booktabs,threeparttable}
\usepackage{multirow}
\draft

\begin{document}
\SetRunningHead{Author(s) in page-head}{Running Head}

\title{Origins of Short Gamma-Ray Bursts Deduced from Offsets to
Their Host Galaxies Revisited}

\author{Xiao-Hong \textsc{Cui} %
}
\affil{School of Physics and State Key Laboratory of Nuclear Physics
and Technology, Peking University, Beijing 100871, China}
\email{xhcui@bac.pku.edu.cn}

\author{Junichi \textsc{Aoi}}
\affil{Yukawa Institute for Theoretical Physics, Oiwake-cho,
Kitashirakawa, Sakyo-ku, Kyoto 606-8502,
Japan}\email{aoi@yukawa.kyoto-u.ac.jp}

\author{Shigehiro {\sc Nagataki}}
\affil{Yukawa Institute for Theoretical Physics, Oiwake-cho,
Kitashirakawa, Sakyo-ku, Kyoto 606-8502,
Japan}\email{nagataki@yukawa.kyoto-u.ac.jp} \and
\author{Ren-Xin {\sc Xu}}
\affil{School of Physics and State Key Laboratory of Nuclear Physics
and Technology, Peking University, Beijing 100871,
China}\email{r.x.xu@pku.edu.cn}

%


\maketitle

\begin{abstract}
The spatial distribution of short Gamma-ray bursts (GRBs) in their
host galaxies provide us an opportunity to investigate their
origins. Based on the currently observed distribution of short GRBs
relative to their host galaxies, we obtain the fraction of the
component that traces the mergers of binary compact objects and the
one that traces star formation rate (such as massive stars) in
early- and late-type host galaxies. We find that the fraction of
massive star component is $0.37\pm0.13$ with error of $1\sigma$
level from the analysis of projected offset distribution. This
suggests that a good fraction of short GRBs still originate from
merger events. From our analysis, we also conclude that the fraction
of late-type hosts among the elliptical, starburst and spiral galaxy
is $0.82\pm0.05$ with error of $1\sigma$ level, which is consistent
with the observed early- to late-type number ratio of host galaxies.
\end{abstract}

\section{Introduction}

After the first discovery of the redshift and the host galaxy
association of the short Gamma-ray burst (GRB with duration
$T_{90}<$ 2s) GRB050509B (Gehrels et al. 2005; Bloom et al. 2006),
observations of the host galaxies of short GRBs (Berger et al. 2005;
Hjorth et al. 2005; Fox et al. 2005; Covino et al. 2006) provide us
an opportunity to study the population of their host galaxies and
the nature of their progenitors (e.g. Prochaska et al. 2006;
Savaglio et al. 2009; Zhang et al. 2009).

The lower explosion energies, the lower star formation rate, lack of
the supernova (SN) association and the locations in their host
galaxies suggest that the mergers of binary compact objects are the
promising progenitor candidates for short GRBs (e.g. Berger et al.
2005; Fox et al. 2005; Belczynski et al. 2006). Based on these
observations, binary compact objects were studied as possible short
GRB progenitors by the population synthesis (PS) methods (e.g.,
Lipunov et al. 1997; Bloom et al. 1999; Fryer et al. 1999;
Belczynski et al. 2002, 2007). By this method, Belczynski et al.
(2006) presented binary compact object formation rates, their merger
rates, locations, and afterglow properties for different initial
conditions. Some predictions from the PS analysis agreed well with
the existing observational constraints such as the redshift
distribution of short GRBs (O'Shaughnessy er al. 2008).

However, not like the long GRBs (GRB with duration $T_{90} > $ 2s)
that are exclusively linked with star-forming galaxies (e.g., Bloom
et al. 1998; Fruchter et al. 1999; Djorgovski et al. 2003;
Christensen et al. 2004; Castro Cer$\acute{o}$n et al. 2006;
Savaglio et al. 2009), short GRBs reside in all types of galaxies
(Berger et al. 2005, 2007; Fox et al. 2005; Gehrels et al. 2005;
Berger 2009). Berger (2009) investigated the host galaxy properties
for all short GRBs localized by Swift X-Ray Telescope (XRT) and
found that the majority of short GRBs appear to occur in
star-forming galaxies. This fact suggests that some short GRBs may
come from other origin that traces the star formation rate. In fact,
Zhang et al. (2009) showed that not only long GRBs, but also a good
fraction of short GRBs could be from the death of the massive stars
(= collapsars). Moreover, Virgili et al. (2009) tried to reproduce
the luminosity-redshift distribution and the peak flux distribution
of short GRBs, and they concluded that the fraction of collapsars as
one of the origins of short GRBs is much higher than binary compact
objects.

In this study, we would like to consider the origin of short GRBs
from the point of view of the spatial distribution of short GRBs in
their host galaxies. It is well known that distribution of binary
compact objects is wider than massive stars, because binary compact
objects can have kick velocities (e.g. Wang, Lai \& Han 2006, Cui et
al. 2006, and references therein) and it takes long time for them to
merge by emitting gravitational waves. We try to obtain the
fractions of the component that traces star formation rate (such as
single massive star component: we call it as ``SMC'' in this study)
and the component of binary compact objects (we call it as ``BC'' in
this study) as the progenitor of short GRBs. We obtain them by
reproducing the observed distribution in their host galaxies with
theoretical models. We use the results of the PS calculations that
give the distribution of BC in late- and early-type host galaxies
(Belczynski et al. 2006). We also use the distribution of SMC in
their host galaxies that is deduced from the observations of
distribution of stars (Bloom et al.  2002). We find that the
fraction of SMC is $0.37\pm0.13$ with error of $1\sigma$ level for
the elliptical, starburst and spiral host galaxies. We also show
that the fraction of the late-type hosts determined by the best
fitting can be consistent with the observed number ratio of late- to
early-type host galaxy. The data and the method are presented in
$\S$ 2. In $\S$ 3, we give the best fitting results for different
type of host galaxy. $\S$ 4 and 5 are our discussions and
conclusion. Throughout the paper, a concordance cosmology with
parameters $H_0 = 71$ km s$^{-1}$ Mpc$^{-1}$, $\Omega_M=0.30$, and
$\Omega_{\Lambda}=0.70$ are adopted.
\section{Data and Method}

As stated above, we try to fit the observed distribution of short
GRBs in their host galaxies by the combination of SMC and BC
distributions. We can obtain the most favored fraction of them from
the fitting (least-square method). In this analysis, we also
investigate the difference of the distribution of short GRBs in
different type of host galaxies. We can obtain the favored fraction
of late-type host galaxies from the fitting (least-square method),
which can be compared with the observed fraction.

\subsection{Observational Data}

In this study, we investigate the distribution of short GRBs in two
ways. One is the distribution of distances of short GRBs from the
center of their host galaxies (strictly speaking, the distane means
the projected distance in the direction perpendicular to the line of
sight). We call it as 'Offset' in this study. Another is same with
Offset, but normalized by the radius of their host galaxies. We call
it as 'Normalized Offset' in this study.

We use 22 samples of short GRBs from previous works. These samples
include the bursts in the original classification (Kouveliotou et
al. 1993) and the ones that have not only a spike of short duration
but also extended emission (EE) as defined by Norris \& Bonnell
(2006). In our sample, 9 bursts have host galaxies with known radius
(e.g. Fong et al. 2010). Therefore, we study the Offset with 22
short GRBs and Normalized Offset with 9 short GRBs. The properties
of the bursts and their host galaxies of our sample are listed in
Table 1. Columns denote the GRB name, the duration ($T_{90}$), the
redshift ($Z$), the Offset ($R_{\rm{projected}}$), the Normalized
Offset ($R_{\rm{projected}}/R_{\rm{e}}$), the type of host galaxy
and the references of these data. Most of the values of $T_{90}$ are
taken from the work of Berger (2009) except for GRB051227,
GRB060505, GRB060614 from Troja et al. (2008) and GRB090426 and
GRB090510 from Swift GRB table
\footnote{http://heasarc.gsfc.nasa.gov/docs/swift/archive/}. We
include three bursts only with the redshift limits of the putative
host galaxies: GRB051210 ($<$1.4), GRB060121 ($>$1.7), and GRB060313
($<$1.1) from the work of Troja et al. (2008) and one burst with the
Offset limit ($<$0.6$''$) for GRB070429B (Cenko et al. 2008). Since
a host galaxy of GBR070809 has not been found to deep limit, the
nearest galaxy is taken as its host galaxy (Fong et al. 2010). The
Normalized Offset of GRB060502B is deduced from the outer component
with a best fit s\'{e}rsic model (Bloom et al. 2007). For the
morphology of the host, elliptical (E) or spiral (S) types were
reported by O'Shaughnessy et al. (2008) and star formation (SF) or
low star formation (LSF) were given by Zhang et al. (2009) and the
reference therein. Others are the bursts without observation reports
(''0'') or too faint (``Faint'') to be observed.
\begin{table*}
 \centering
 \begin{minipage}{140mm}
  \caption{Properties of the short GRB samples}
  \begin{tabular}{@{}llrrrrlrlr@{}}
  \hline
    GRB & $T_{90}$ & Z & $R_{\rm{projected}}$ & $R_{\rm{projected}}/R_{\rm{e}}$ & Type& Ref  \\
& (s) & & (kpc)&\\
 \hline
 050509B &   0.04    &   0.225   &   54.3 (12.1)   &   2.59    $\pm$   0.58  & E  & 1,2,9\\
050709*   &   0.07    &   0.161   &   3.70 (0.03)    &   2.04    $\pm$   0.02   & S & 1,2,9\\
050724*   &   3   &   0.258   &   2.69 (0.07)    &   1.28    $\pm$   0.05   & E & 1,2,9\\
051210   &   1.27    &   $<$1.4 &   30.3 (19.5)    &   5.66    $\pm$   3.65  & 0 & 1,2 \\
051221A &   1.4 &   0.546   &   2.05 (0.19)    &   0.88    $\pm$   0.08 & S   & 1,2,9 \\
060121   &   1.97    &   $>$1.7 &   0.96 (0.37)    &   0.18    $\pm$   0.07  & Faint & 1,2  \\
060313   &   0.7 &   $<$1.1 &   2.57 (0.53)    &   1.66    $\pm$   0.32  & Faint  & 1,2\\
060502B &   0.09    &   0.287   &   70.0 (16.0)  &  6.66 $\pm$ 1.52 & E &2,8,9\\
060505   &   4   &   0.089   &   7.45 (0.53)    &   ...  & S &2,9\\
060614*   &   103 &   0.125   &   $\sim$1.10   &   ... & LSF &2,10\\
060801   &   0.5 &   1.131   &   19.7 (19.8)    &   ... & 0  & 2\\
061006*   &   0.42    &   0.438   &   1.37 (0.27)    &   0.41    $\pm$   0.09 & LSF &1,2,10   \\
061201   &   0.8 &   0.111   &   33.9 (0.40) &   ...  & SF  &2,10      \\
061210*   &   0.19    &   0.41    &   10.7 (9.70) &   ... & SF &2,10       \\
061217   &   0.21    &   0.827   &   55.0 (28.0)  &   ...   & SF    &2,10  \\
070429B &   0.5 &   0.9023  &   $\sim$16.99    &   ...    & Faint   & 3,10  \\
070714B* &   3   &   0.9225  &   $\sim$11.64     &   ...   & S  & 4,10    \\
070724A &   0.4 &   0.457   &   4.80 (0.10) &   ...    & SF  &2,10   \\
070809   &   1.3 &   0.2187  &   $\sim$20.0   &   ...       & S  & 1,10\\
071227   &   1.8 &   0.381   &   15.0 (2.20) &   ...  & S  &5,10     \\
090426   &   1.2 &   2.609   &   $\sim$0.80         &   ...   & SF  &6,10    \\
090510   &   0.3 &   0.903   &   $\sim$5.50        &   ...   & SF   &7,10   \\
\hline figure\end{tabular}
\begin{tablenotes}
  \item[*]* Burst with extended emission.
 \end{tablenotes}
 \begin{tablenotes}
  \item[*]References: (1) Fong et al. 2010; (2) Troja et al. 2008; (3) Cenko et al. 2008;
  (4) Graham et al. 2009; (5) D'Avanzo et al. 2009;
  (6) Levesque et al. 2009; (7) Rau et al. 2009; (8) Bloom et al.
  2007; (9) O'Shaughnessy et al. 2008; (10) Zhang et al. 2009
 \end{tablenotes}
\end{minipage}
\end{table*}

From Table 1, we can see that the short GRB host galaxies include
not only early-type but also late-type. The observed number ratio of
late- to early-type host galaxy is 5:1 (Berger 2009; Fong et al.
2010).

\subsection{Method of Analysis}

Belczynski et al. (2006) developed an updated PS code to calculate
the locations of mergers of binary compact objects in early- and
late-types host galaxies: elliptical (Ellip), spiral (Sp), and
starburst (SB) galaxy. When they calculated the motion of the
binaries, they took into account the gravitational potential of
their host galaxies. In their study, the mass density of their host
galaxies are assumed to be constant, and they modeled gravitational
potentials for each types of galaxies with a large and a small mass.
The bulge mass and radius for the large elliptical galaxy are taken
as $5\times 10^{11}M_{\odot}$ and 5kpc, respectively. The total mass
of the bulge and disk and the disk radius for the large spiral and
the starburst galaxies are assumed as $10^{11}M_{\odot}$ and 12kpc.
The small galaxies are downscaled by a factor of $10^3$ in mass and
of 10 in size (constant density). The Offset of NS-NS (neutron star
and neutron star, defined as ``NN'' in this work) and NS-BH (neutron
star and black hole, ``NB'') merger locations were given by
Belczynski et al. (2006) for different types of host galaxies.
Although Belczynski et al. (2002) presented all types of the binary,
in this work we consider two types of the binary: NN and NB in three
kinds of host galaxies: Ellip, Sp, and SB as mentioned by Belczynski
et al. (2006). The number ratio of the NN to NB merger components
has been calculated in Belczynski et al. (2002). They evolved
$N_{\rm{tot}}=3\times 10^7$ initial binaries in a spiral galaxy and
found that 52,599 NN and 8,105 NB are formed, which implies that the
ratio of NN to NB is 6.49. They also found that the evolutionary
time (the time required for the initial progenitor binary on ZAMS to
form a binary) are typically of the order of a few to several tens
Myr, and their distributions are similar among different types of
galaxies. Thus it will be reasonable to assume this ratio to be the
same for every galaxy since the structure of the star cluster and
the evolution of the binary in them will be the similar in every
type of the galaxy. Therefore, here we fix this
parameter as 6.49 for each type of galaxy. 

As mentioned by Smith et al. (2005) and Belczynski et al. (2006),
most elliptical galaxies are formed before z $\sim$ 2 and there are
no longer star-forming regions, while late-type galaxies have an
ongoing and active star-forming regions. This suggests that in the
spirals and starbursts galaxies, GRBs may come not only from BC but
also from SMC. Thus we consider SMC in this study. As for the model
of SMC, Bloom et al. (2002) proposed a model of the number density
of massive-star forming regions in a disk galaxy as
\begin{equation}
N(r)dr\propto  r exp(-1.67r)dr, \label{k}
\end{equation}
where $r=R/R_{\rm{half}}$, the half-light radius of a galaxy
$R_{\rm{half}}=1.67\times R_{\rm {e}}$ if $R_{\rm {e}}$ is the disk
scale length. It is true that the distribution of massive star
forming regions in a galaxy has large uncertainty as mentioned in
their work, this distribution function is frequently used in the
analysis of Offset/Normalized Offset of GRBs. Thus we use this
distribution as a template in this study.

In summary, in our study, we consider only the distributions of NN
and NB (that is, BC) for early-type host galaxies, while we consider
distributions of BC and SMC for late-type host galaxies. The
procedure of analysis includes the following steps. (1) We derive
theoretical Offset/Normalized Offset curves for each type of host
galaxy with a help of the PS model and SMC model. It is noted that a
single theoretical curve can not obtained since there are
uncertainties in the models. Thus we obtained one hundred curves for
an each host galaxy changing the values for uncertain parameters.
(2) Since the host galaxies for short GRBs include all types, we
combine two/three distributions of Offset/Normalized Offset for
different types of host galaxies with proper weight. When we
consider the contribution from the late-type galaxies, we add the
SMC component with proper weight. The proper weights are determined
by the the least-square method to reproduce the observed
Offset/Normalized Offset. (3) We apply the statistic methods to test
all the fittings and give the test results.

We give an example below. Let's consider Ellip-SB-Sp model. In this
model, there are five components: two SMCs and two BCs in late-type
host galaxies (Sp-SB), and one BC in early-type host galaxy (Ellip).
We introduce the fractions $fr_1$ and $fr_2$ that are the fractions
of the SMC in SB and Sp galaxies respectively, and $frG_1$ and
$frG_2$ that are the fractions of SB and Sp galaxies in all types of
galaxies. The most favored Offset/Normalized Offset curves are
obtained by the least-square method for
\begin{eqnarray}
P_{\rm{fit}}=&[P_{\rm{SMC,1}}\times fr_1+P_{\rm{Merg,1}}\times
(1-fr_1)] \times frG_1\nonumber \\
&+[P_{\rm{SMC,2}}\times fr_2+P_{\rm{Merg,2}}\times (1-fr_2)] \times frG_2\nonumber \\
&+P_{\rm{Merg,3}}\times (1-frG_1-frG_2), \label{fraction}
\end{eqnarray}
where $P_{\rm{SMC,1}}$ and $P_{\rm{Merg,1}}$ are the Offset curves
of SMC and BC in SB galaxy. $P_{\rm{SMC,2}}$ and $P_{\rm{Merg,2}}$
are those in Sp galaxy, and $P_{\rm{Merg,3}}$ is that of BC in Ellip
galaxy. The fractions of SMC and of the late-type host galaxy are
$fr_{\rm{SMC}}=fr_1 \times frG_1+fr_2 \times frG_2$ and
$fr_{\rm{late-type}}=frG_1+frG_2$, respectively. For Ellip-Sp model,
$P_{\rm{SMC,1}}=P_{\rm{Merg,1}}=0$ while for Ellip-SB model,
$P_{\rm{SMC,2}}=P_{\rm{Merg,2}}=0$. In summary, we obtain the proper
weights $fr_1$, $fr_2$, $frG_1$, and $frG_2$ by reproducing the
observed Offset/Normalized Offset using the theoretical curves
$P_{\rm{SMC,1}}$, $P_{\rm{Merg,1}}$ , $P_{\rm{SMC,2}}$,
$P_{\rm{Merg,2}}$, and $P_{\rm{Merg,3}}$.

\section{Results}

The results for the Offset are shown in Figure 1. It is shown at the
top what kind of host galaxies are considered in each panel.
Horizontal axis represents the projected distance from the center of
a host galaxy, while vertical axis represents the cumulative Offset.
Red circles represent observed 22 samples. The blue dashed curve is
the best fitting curve. Green solid curve is the contribution from
SMC, and blue solid curve is the one from BC in the best fitting
case. The deep/bright gray regions in the left panels represent
1$\sigma$/3$\sigma$ error ranges.  In the right panels, all of the
possible curves are shown for SMC and BC. For example, at the left
bottom panel of Figure 1, the result for Ellip-SB-Sp model is shown
where BC is expressed as $P_{\rm{Merg,1}}\times (1-fr_1) \times
frG_1+P_{\rm{Merg,2}}\times (1-fr_2) \times
frG_2+P_{\rm{Merg,3}}\times (1-frG_1-frG_2)$. From Figure 1, we can
find that the observed data are all in the 3$\sigma$ error ranges of
best fitting results.
\begin{figure*}
\includegraphics[angle=0,scale=0.50]{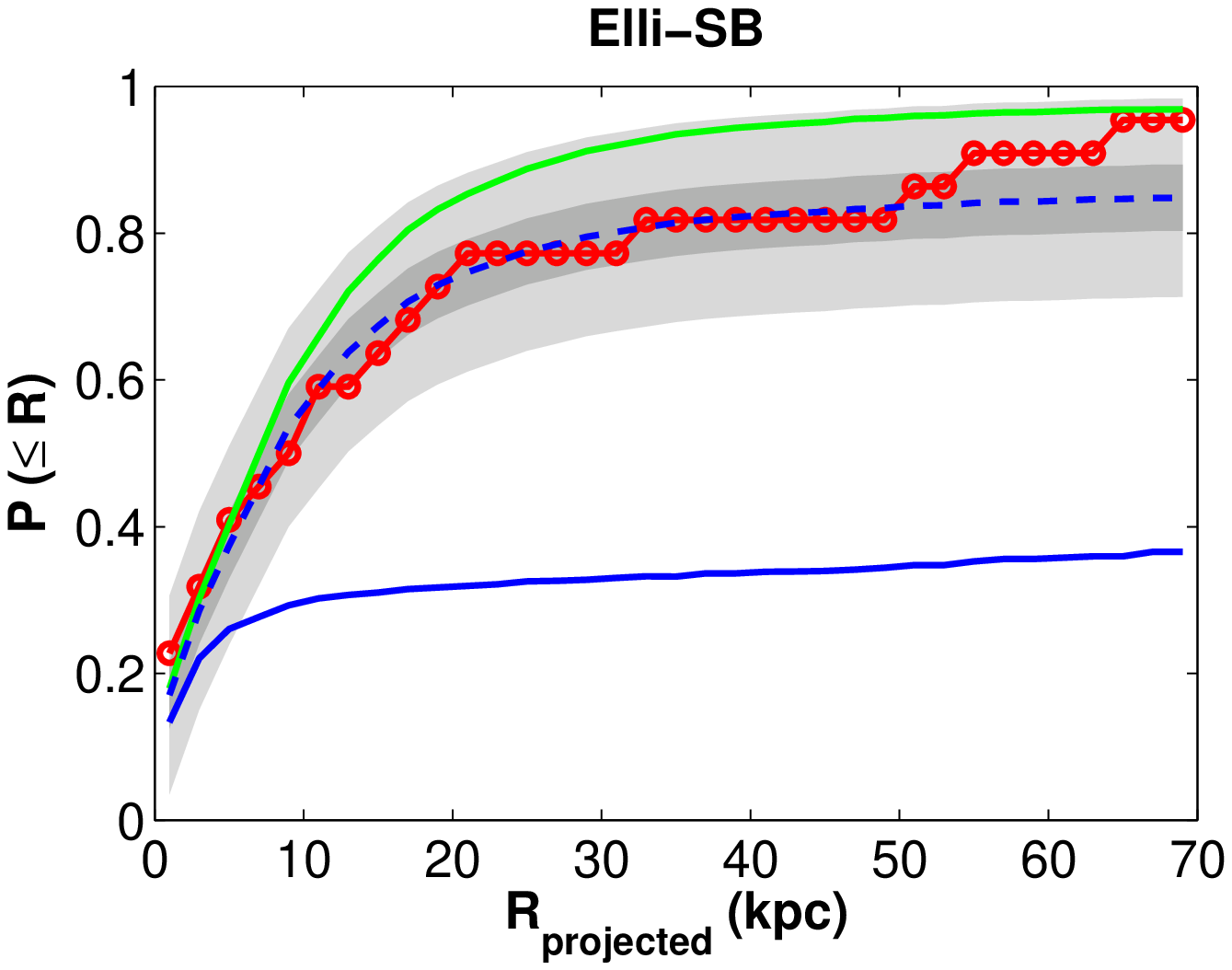}
\includegraphics[angle=0,scale=0.50]{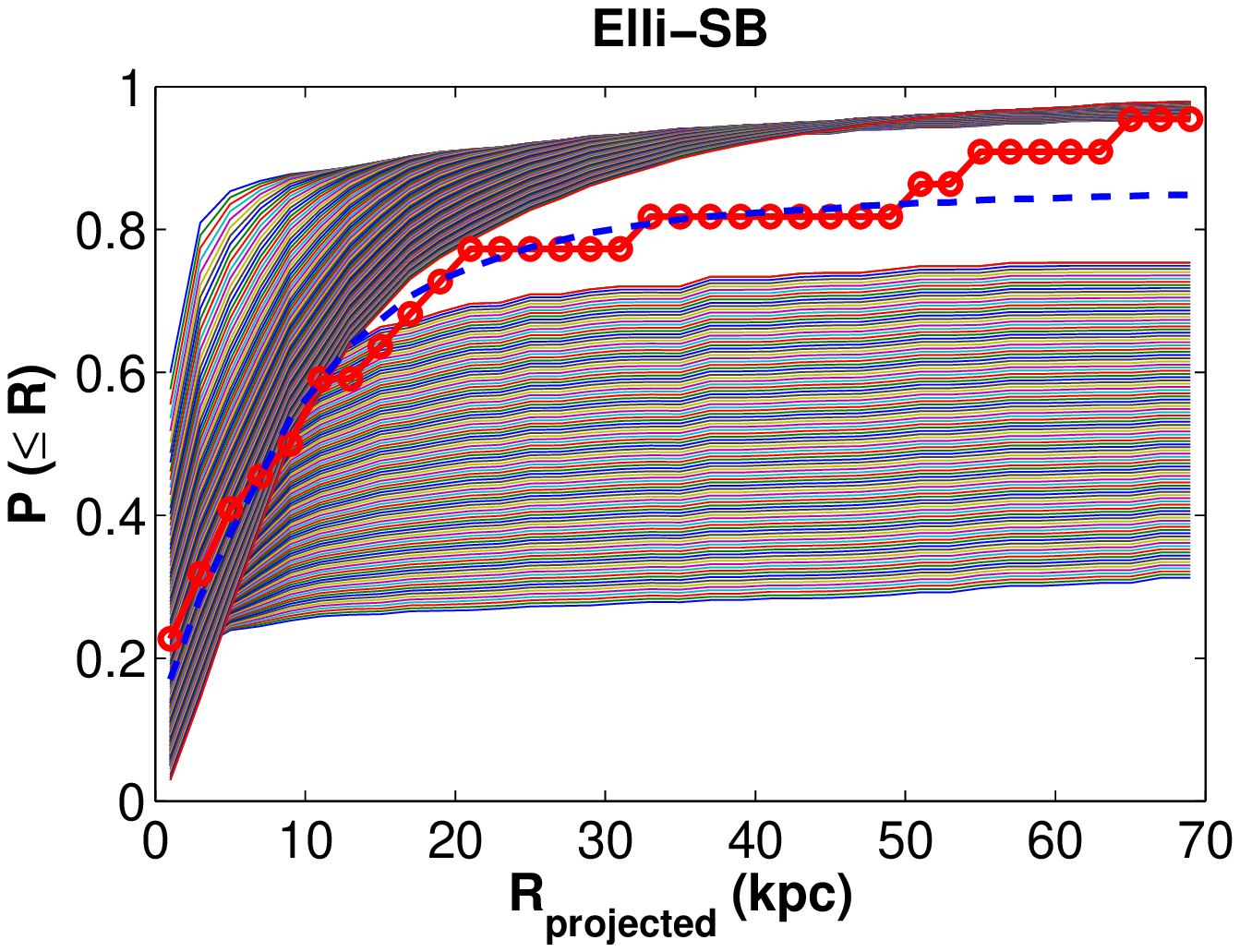}
\includegraphics[angle=0,scale=0.50]{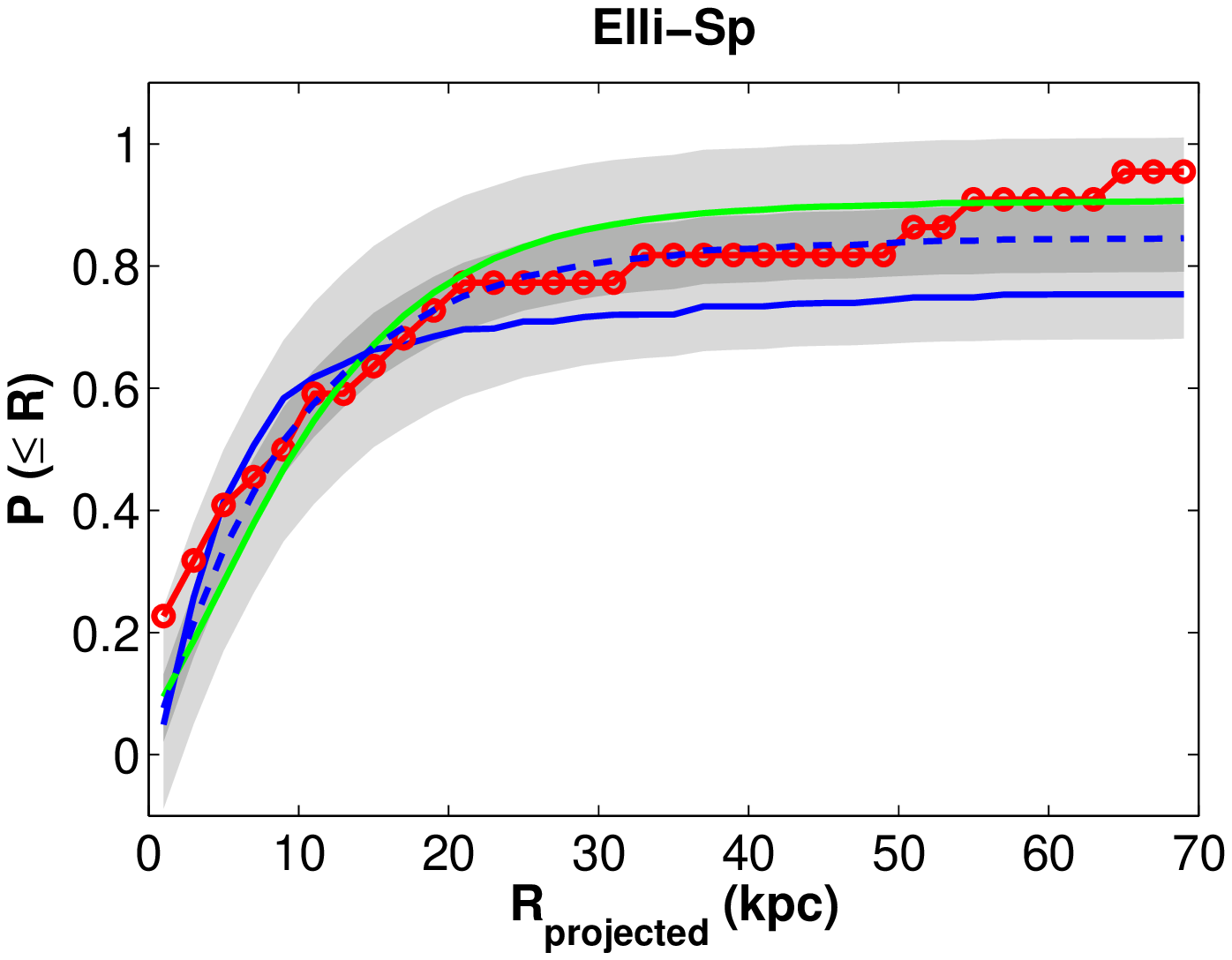}
\includegraphics[angle=0,scale=0.50]{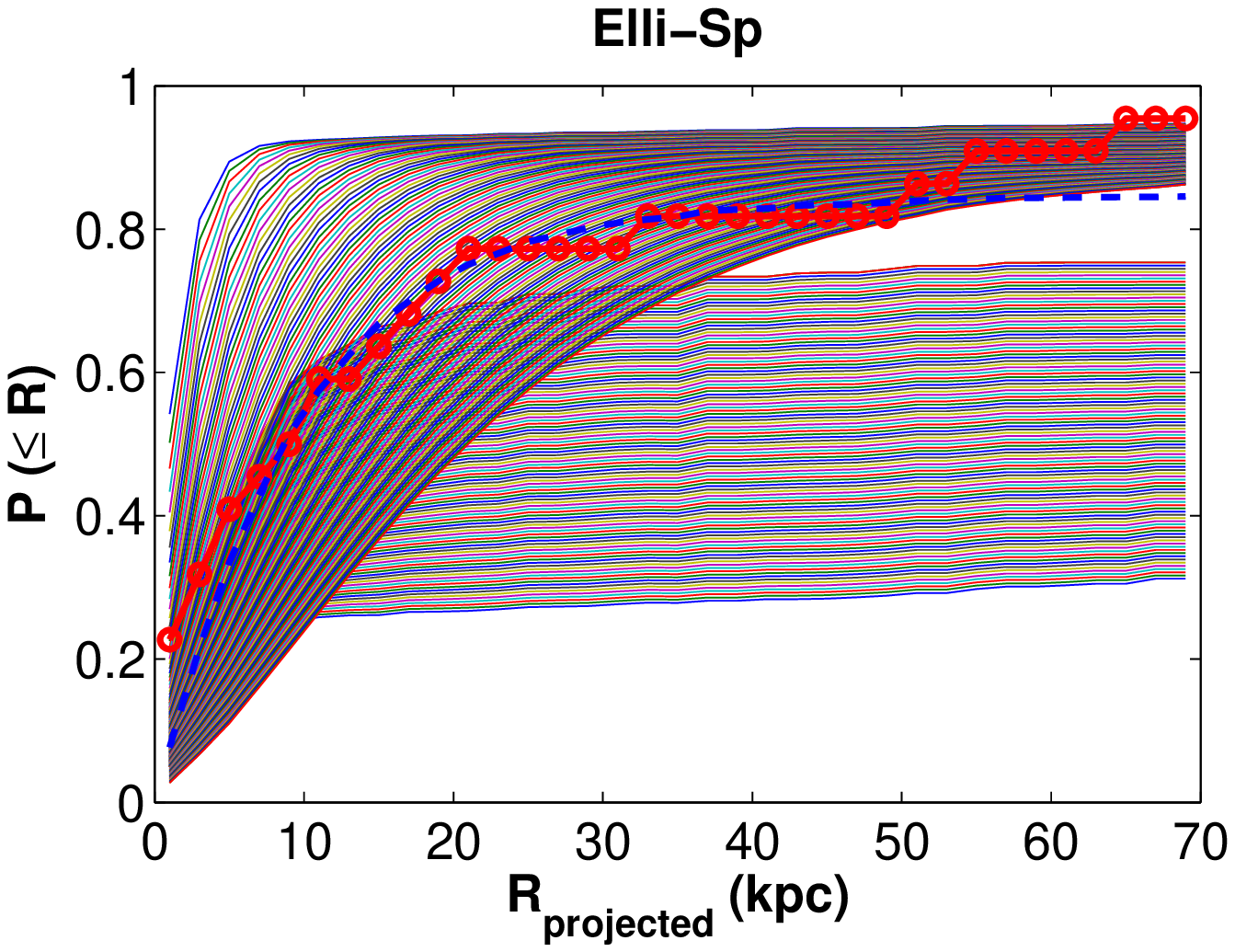}
\includegraphics[angle=0,scale=0.50]{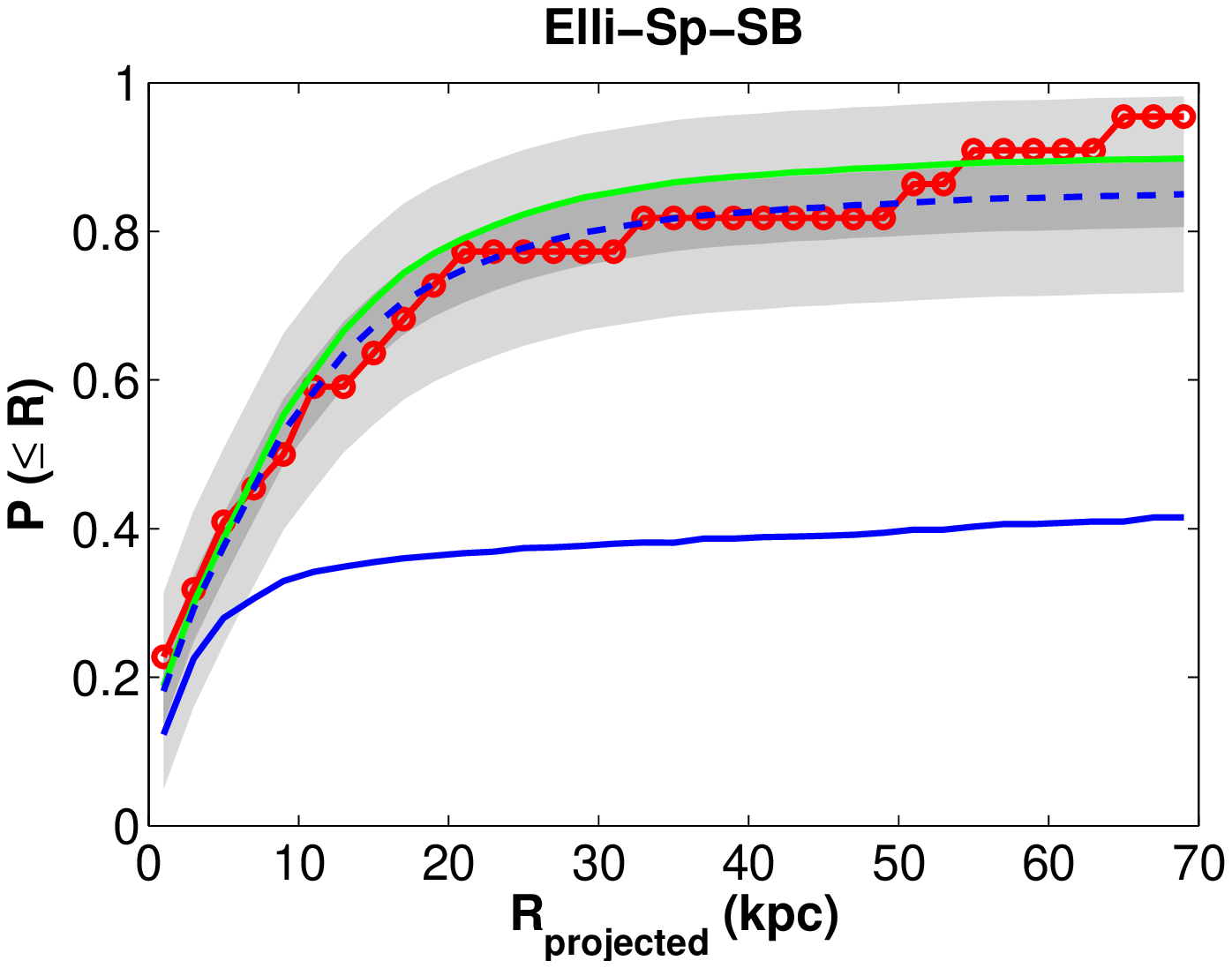}
\includegraphics[angle=0,scale=0.50]{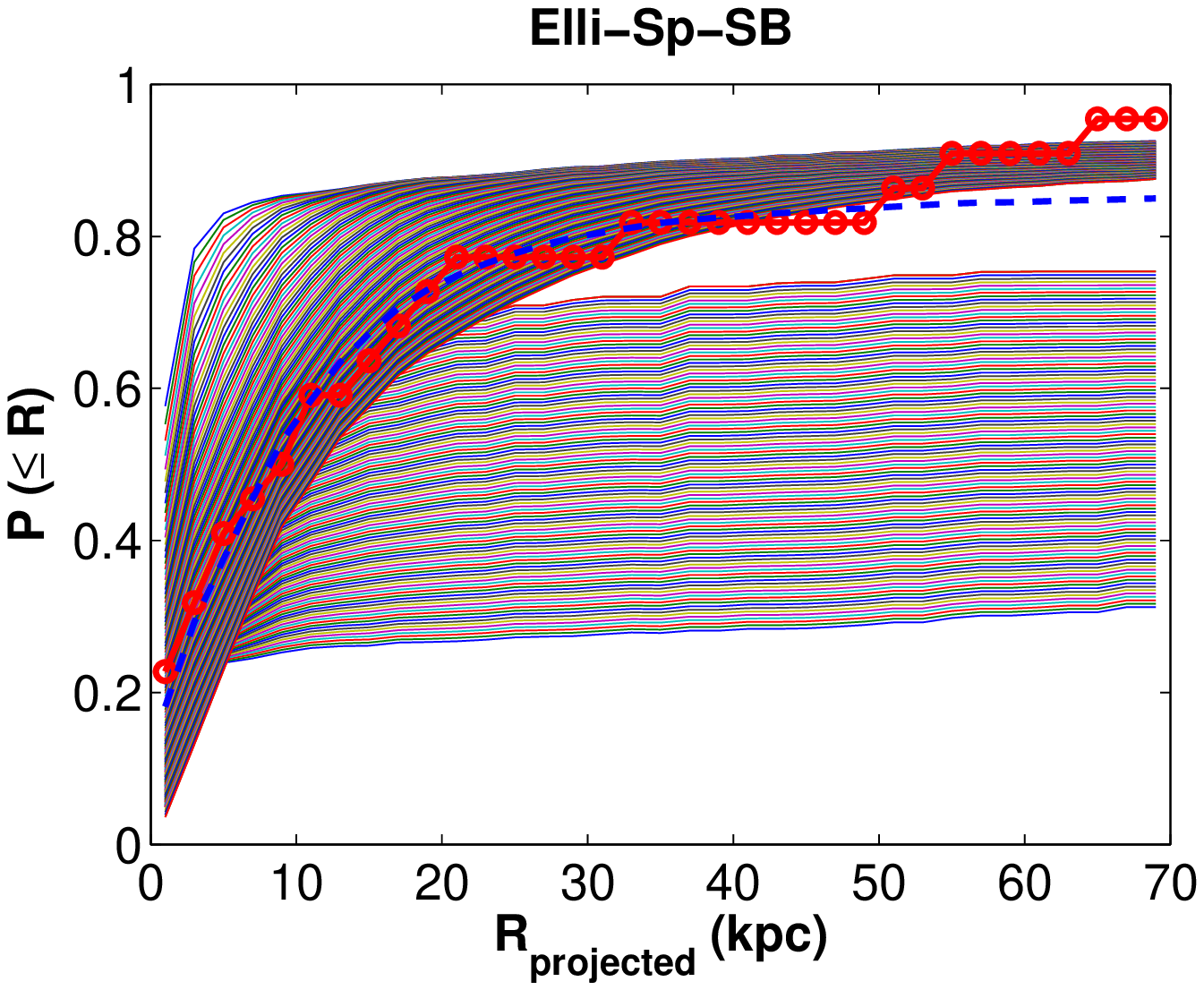} \hfill
\caption{ Results for the cumulative Offset distribution. The top
label represents what kind of host galaxies are considered.
Horizontal axis represents the projected distance from the center of
a host galaxy, while vertical axis represents the cumulative Offset.
Red circles represent observed 22 samples. The blue dashed curve is
the best fitting curve. Green solid curve is the contribution from
SMC, and blue solid curve is the one from BC in the best fitting
case. The deep/bright gray regions in the left panels represent
1$\sigma$/3$\sigma$ error ranges. In the right panels, all of the
possible curves are shown for SMC and BC. }
\label{offset-distribution1}
\end{figure*}

The values of $\chi^2_{\rm{min}}$, reduced masses of the early- and
late-type galaxies, fraction of SMC component in late-type galaxy
and its error, fraction of the late-type host galaxy and its error,
the significant level (p) and statistic results (ksstat, tstat,
fstat) for K-S test, T-test and F-test to the best fitting curve are
shown in Table 2. From the table, we find that the fractions of the
late-type host galaxy and the SMC components are 0.82$\pm$0.05 and
0.37$\pm$0.13 with error of $1\sigma$ level for the Offset analysis.
Here we would like to explain how we calculate the ``reduced'' mass
of host galaxies. They are expressed as $M_{\rm{Merg,3}}$,
$M_{\rm{SMC,1}}\times fr_1+M_{\rm{Merg,1}}\times (1-fr_1)$, and
$M_{\rm{SMC,2}}\times fr_2+M_{\rm{Merg,2}}\times (1-fr_2)$ for
elliptical, starburst, and spiral galaxies. Here $M_{\rm{SMC,1}}$ is
the mass of SB galaxy that gives the probability distribution
$P_{\rm{SMC,1}}$, and $M_{\rm{Merg,1}}$ is the one that gives
$P_{\rm{Merg,1}}$, and so on.
\begin{table}
\caption[]{Values of $\chi^2_{\rm{min}}$, the masses of the host
galaxies, the fraction of the late-type host galaxies with $1\sigma$
error in parentheses, the faction of SMC in parentheses, and the
test results for the best fitting of Offset and Normalized Offset
(last three columns) distribution.} \label{offset fixed} \centering
\begin{tabular}{|c|c|c|c|c|c|c|c|c|} \hline \hline
\multicolumn{2}{|c|}{sample}& \multicolumn{3}{|c|}{Offset Analysis}
&
\multicolumn{3}{|c|}{Normalized Offset Analysis} \\
\cline{1-8}
\multicolumn{2}{|c|}{early-type galaxy}& \multicolumn{3}{|c|}{Ellip} & \multicolumn{3}{|c|}{Ellip} \\
\cline{1-8}
\multicolumn{2}{|c|}{late-type galaxy}             & SB   &  Sp   & SB-SP & SB   &  Sp   & SB-SP \\
\hline
\multicolumn{2}{|c|}{$\chi^2_{\rm{min}}$}               & 0.20 & 0.24  & 0.19 & 0.28  & 0.37&  0.28   \\
\hline
\multicolumn{2}{|c|}{$M_{\rm{early}}(10^{10}M_{\odot})$}& 6.97 & 42.0 &12.4 & 50& 50  & 50  \\
\hline
\multicolumn{2}{|c|}{$M_{\rm{late}}(10^{10}M_{\odot})$} & 6.04 & 4.86 &5.55 & 10 & 0.21    &10  \\
\hline
\multirow{2}{*}{fraction}   & late-type       & 0.83 (0.10) & 0.74 (0.05) &0.82 (0.05) & 0.31 (0.27)  &0.24 (0.30)    &0.30 (0.21)   \\
\cline{2-8}
& SMC component                                       & 0.32 (0.13) & 0.48 (0.14) &0.37 (0.13) & 0.17 (0.23) & 0.24 (0.41)   & 0.19 (0.33)  \\
\hline
\multirow{2}{*}{KS test} 
& p                                                & 0.09 & 0.09 &0.09 & 0.02 & 0.02 &0.02     \\
\cline{2-8}
& ksstat                                           & 0.29 & 0.29 & 0.29 & 0.58 &0.58   &0.58     \\
\hline
\multirow{2}{*}{T-test}  
& p                                                & 0.68 & 0.60 &0.71  & 0.70 & 0.76   & 0.71 \\
\cline{2-8}
& tstat                                            & 0.41 & 0.52 &0.37 & -0.39& 0.31   & -0.37  \\
\hline
\multirow{2}{*}{F-test}                     
& p                                                & 0.76 & 0.73 &0.75  & 0.45  & 0.09  & 0.41  \\
\cline{2-8}
& fstat                                            & 1.11 & 0.89 &1.12  & 1.60  &  2.88   & 1.67   \\
\hline\hline
\end{tabular}
\end{table}

Then from Table 2, we find that the fraction ($0.82\pm0.05$ with
error of $1\sigma$ level) of the late-type host galaxy in
Ellip-SB-Sp model is consistent with the observed number ratio (5:1)
(Berger 2009; Fong et al. 2010). It should be noted that the values
of $\chi^2_{\rm{min}}$ and the test results for the Ellip-Sp-SB
model are the best among all models, because short GRBs are found in
every type of galaxies.

The results of Normalized Offset are shown in Figure 2 and Table 2.
The fractions of the late-type host galaxies and those of SMC are
also presented there. The GRB identifications are noted along the
solid histogram in Figure 2. We can see that the observed data are
all in the 2$\sigma$ error ranges of the best fitting results. We
also find that the fractions of SMC for all types of the host
galaxies (Ellip-Sp-SB) and the late-type host galaxies are
0.19$\pm$0.33 and 0.30$\pm$0.21, respectively. The
$\chi^2_{\rm{min}}$ and the errors of the fractions for Normalized
Offset are larger than the Offset analysis. This is likely because
the number of Normalized Offset sample is smaller. Thus, the results
for the Offset analysis will be more reliable than Normalized Offset
in this work. However, with new observations for the short GRB host
galaxies in the future, the Normalized Offset analysis should shed
light on the origin of short GRBs.
\begin{figure*}
\includegraphics[angle=0,scale=0.50]{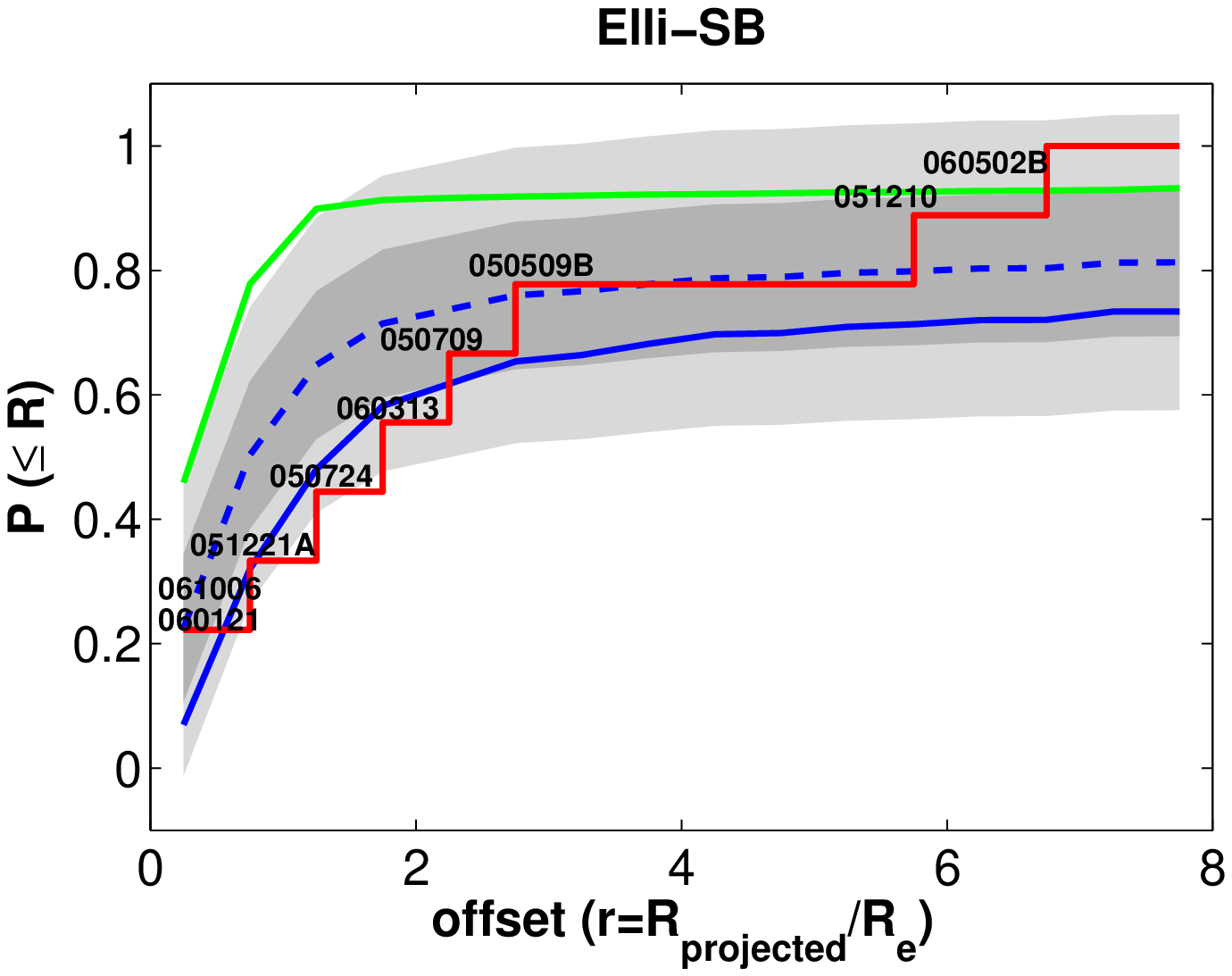}
\includegraphics[angle=0,scale=0.50]{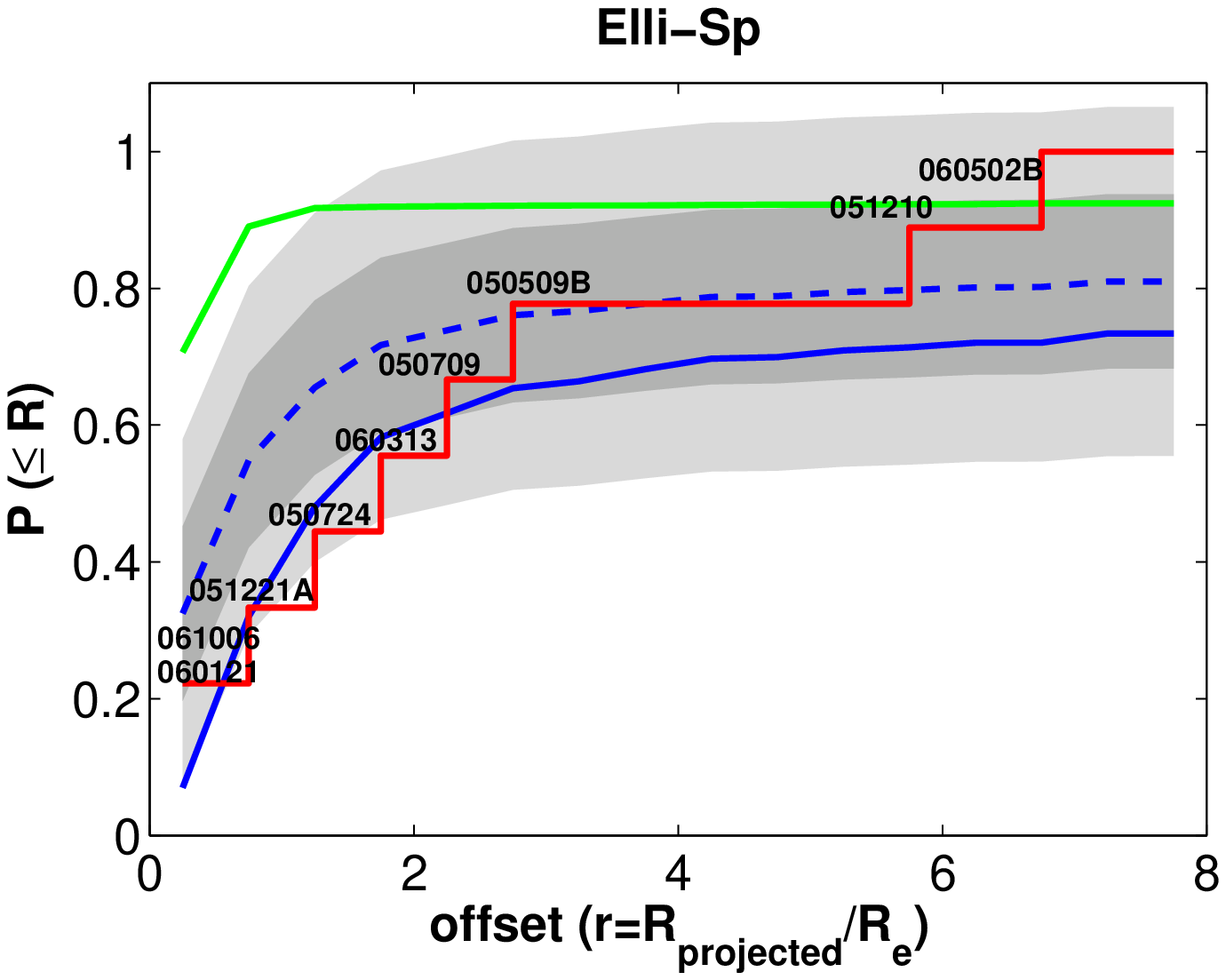}
\includegraphics[angle=0,scale=0.50]{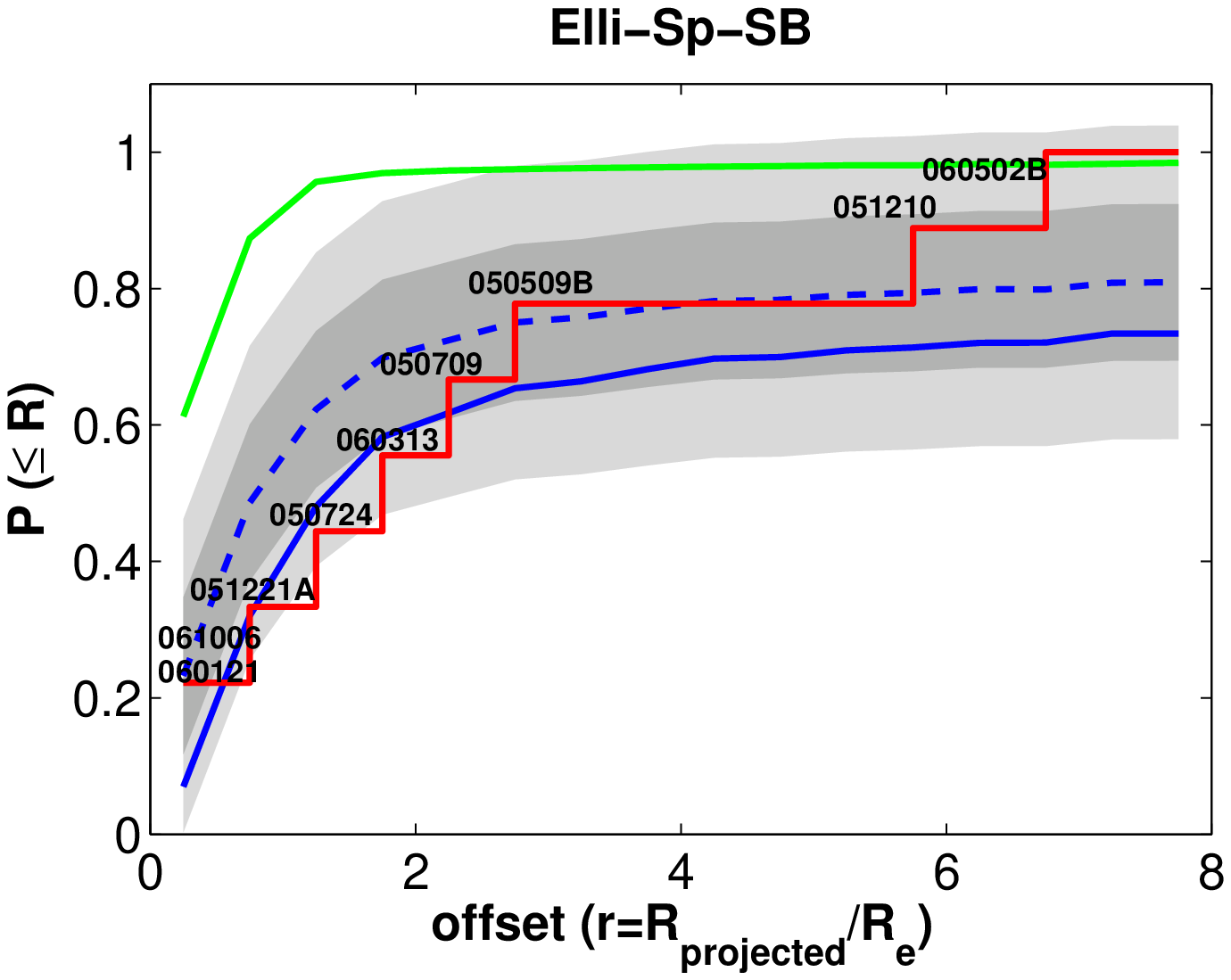}
\includegraphics[angle=0,scale=0.50]{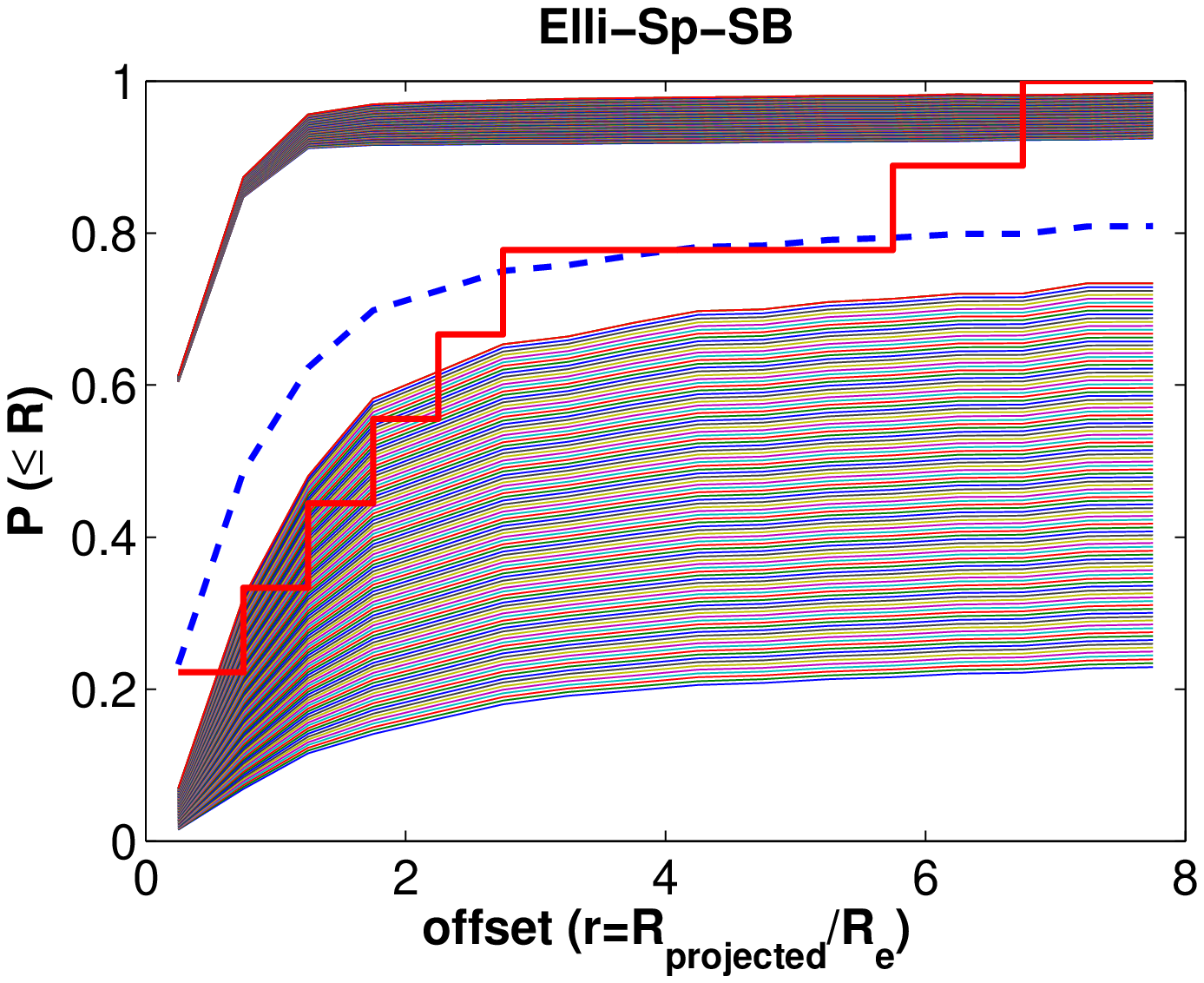}
\hfill \caption{ Same with Figure 1, but for results for the
Normalized Offset. Red line is the observed data. The GRB
identifications are noted along the red line. The deep/bright gray
regions represent 1$\sigma$/2$\sigma$ error ranges. In lower right
panel, all of the possible curves are shown for SMC and BC for the
Ellip-Sp-SB model.}
\end{figure*}

Figure 3 shows the contour plots of $\chi^2$ in the plane of the
fractions of SMC and late-type galaxies for the Offset (left panel)
and Normalized Offset (right panel) analysis.The Ellip-SB-Sp model
is adopted. The minimum of $\chi_{\rm{min}}^2$ is denoted by the red
star. Other parameters are fixed to be the most favored values by
the least-square method.
\begin{figure*}
\includegraphics[angle=0,scale=0.40]{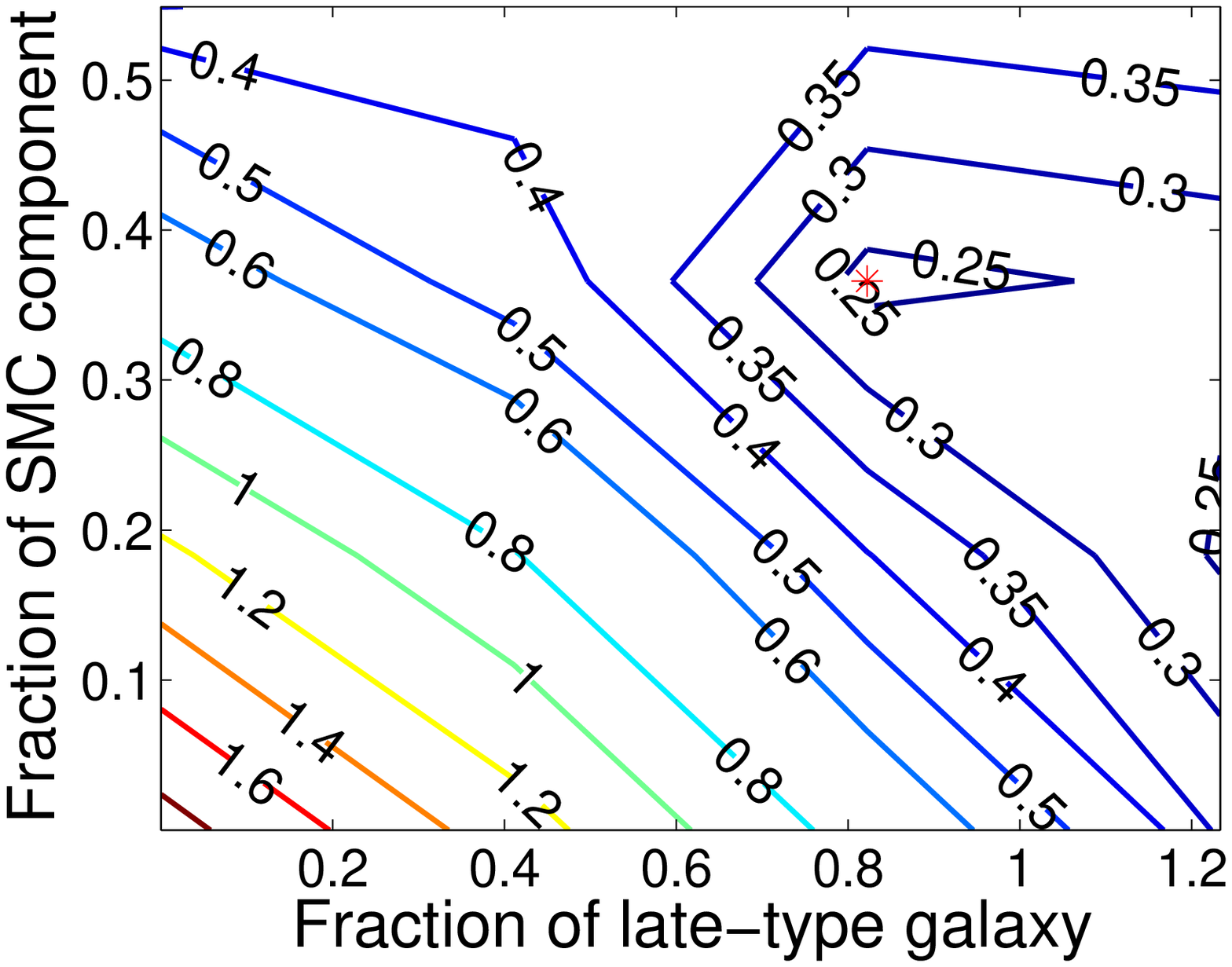}
\includegraphics[angle=0,scale=0.40]{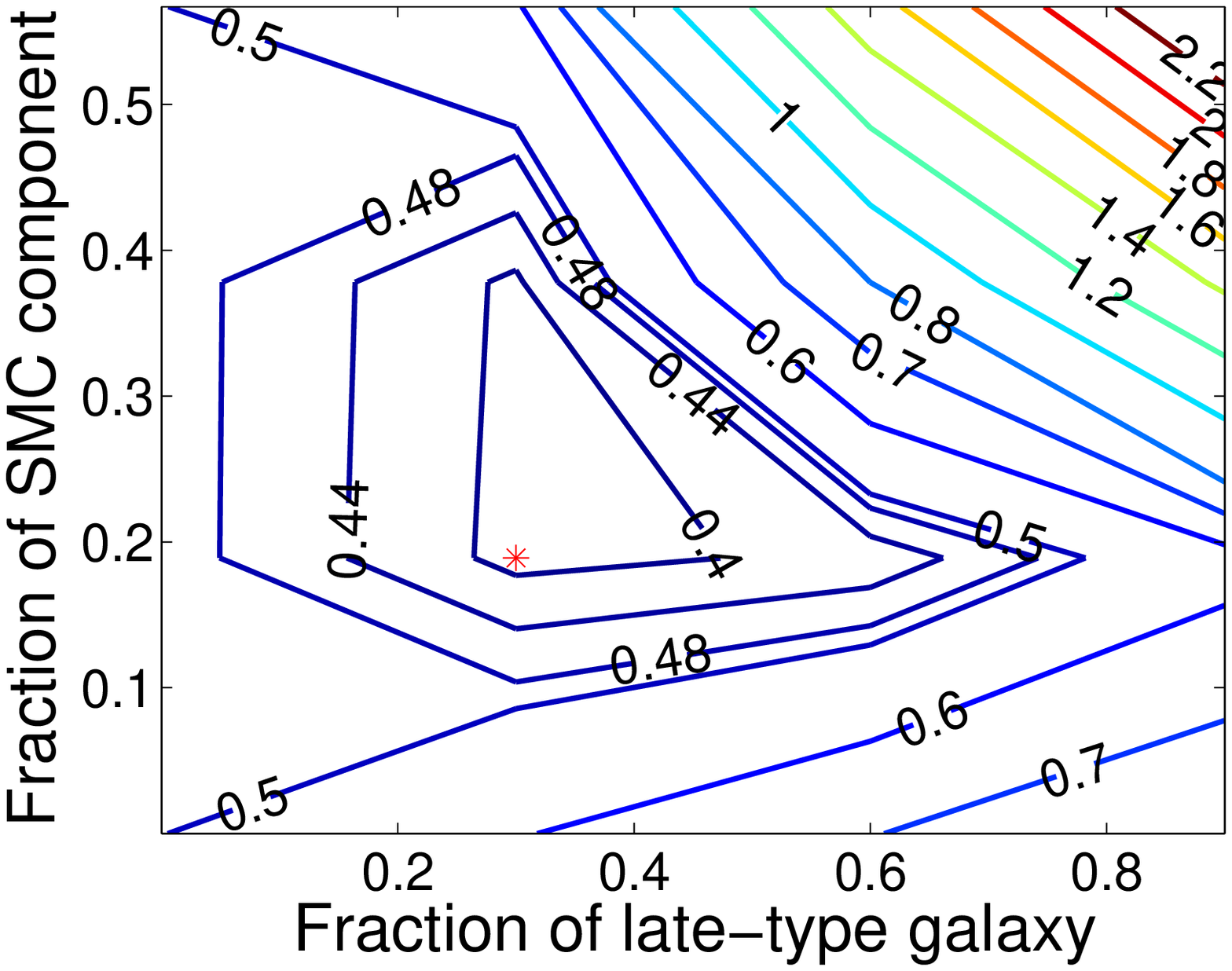}
\hfill \caption{Contour plot of $\chi^2$ in the plane of the
fractions of SMC and late-type galaxies. The Ellip-SB-Sp model is
adopted. The minimum of $\chi_{\rm{min}}^2$ is denoted by the red
star. Left panel: the Offset analysis. Right panel: the Normalized
Offset analysis. } \label{contour}
\end{figure*}

\section{Discussion}

Troja et al. (2008) analyzed the different properties of short GRB
with extended emission (EE) and those without EE and found that the
bursts with large offsets have no observed EE components. Here we
also investigate the Offset for these two sub-samples in our
observed offset sample: one includes 6 bursts with EE and the other
is composed of 16 bursts without EE. The results are summarized in
Table 3. The fitting results show that the average fractions of the
late-type galaxies are about 0.58 and 0.62 for the bursts without EE
and for those with EE. The average SMC component fractions for two
samples are also very near: $\sim$0.48 and $\sim$0.52. Taking into
account the error bars, we could not find any difference between the
bursts with and those without EE in this analysis.
\begin{table}
\caption[]{Same with Table 2, but the Offset analysis for the bursts
without/with EE. } \label{EE} \centering
\begin{tabular}{|c|c|c|c|c|c|c|c|c|c|c|c|} \hline \hline
\multicolumn{2}{|c|}{sample}& \multicolumn{3}{|c|}{Offset Analysis
for
the bursts without EE} & \multicolumn{3}{|c|}{Offset Analysis for the bursts with EE} \\
\cline{1-8}
\multicolumn{2}{|c|}{early-type galaxy}& \multicolumn{3}{|c|}{Ellip} & \multicolumn{3}{|c|}{Ellip} \\
\cline{1-8}
\multicolumn{2}{|c|}{late-type galaxy}                  &SB  & Sp& SB-Sp &SB  & Sp &  SB-Sp   \\
\hline
\multicolumn{2}{|c|}{$\chi^2_{\rm{min}}$}               & 0.30   & 0.35& 0.31& 0.20 & 0.19     & 0.19            \\
\hline
\multicolumn{2}{|c|}{$M_{\rm{early}}(10^{10}M_{\odot})$}& 49.0 &50& 50 & 37.6  & 22.8   & 7.96           \\
\hline
\multicolumn{2}{|c|}{$M_{\rm{late}}(10^{10}M_{\odot})$} & 10    &7.43&10  & 0.21 & 0.31   &0.41           \\
\hline
\multirow{2}{*}{fraction}   & late-type                 & 0.61 (0.09)  &0.54 (0.11)& 0.60 (0.08)  & 0.67 (0.18) &0.58 (0.15)   &0.65 (0.17)                \\
\cline{2-8}
& SMC component                                         & 0.48 (0.12)  &0.53 (0.11) &0.43 (0.06) & 0.45 (0.20)  & 0.54 (0.20)  & 0.57 (0.20)             \\
\hline
\multirow{3}{*}{KS test} 
& p                                                 & 0.23  &0.23&0.04 & 0.43   & 0.43    &0.43       \\
\cline{2-8}
& ksstat                                           & 0.23  & 0.23& 0.30 & 0.33  &0.33     &0.33      \\
\hline
\multirow{3}{*}{T-test}                        
& p                                                & 0.88  &0.87 &0.82 & 0.65  & 0.78   & 0.82        \\
\cline{2-8}
& tstat                                            & 0.15  &0.16& 0.23 & 0.47   & 0.28    & 0.23     \\
\hline
\multirow{3}{*}{F-test}                      
& p                                                & 0.96  &0.95&0.78 & 0.54   & 0.50   & 0.39      \\
\cline{2-8}
& fstat                                            & 1.02 &1.02&1.09  & 1.46    &  1.51   & 1.70      \\
\hline\hline
\end{tabular}
\end{table}

We check the correlation between the Offset and Normalized Offset.
In Figure 4, we show the observed correlation and theoretically
calculated one. The region between solid lines are expected range
for early-type host galaxies in our calculation. The one between
green lines are for late-type galaxies. The boundaries of these
regions correspond to the maximum and minimum mass of the host
galaxies in our analysis. Red points denote that the host galaxies
are early-type, while black points denote that the ones are
late-type. Except for GRB050509B with large host galaxy radius, all
the observed data are in the expected range by our models.
Especially for the late-type host galaxy, they are all in the
expected range. They seem to have two different groups: one is
GRB050509B, GRB060502B and GRB051210 with larger Offsets and
Normalized Offsets. Two of their host galaxies are elliptical, but
the one for GRB051210 is too faint to determine its type. It will be
interesting if more bursts in this group will be found in the future
and if their host galaxies are exclusively elliptical. The others
including bursts with EE seem to trace a linear distribution. The
linear correlation coefficient is $R\simeq0.99$. There seems to be
no significant difference for the bursts with and without EE. We
also find $R\simeq0.74$ for all the observed data. These discussions
are consistent with Troja et al. (2008).
\begin{figure*}
\includegraphics[angle=0,scale=1.00]{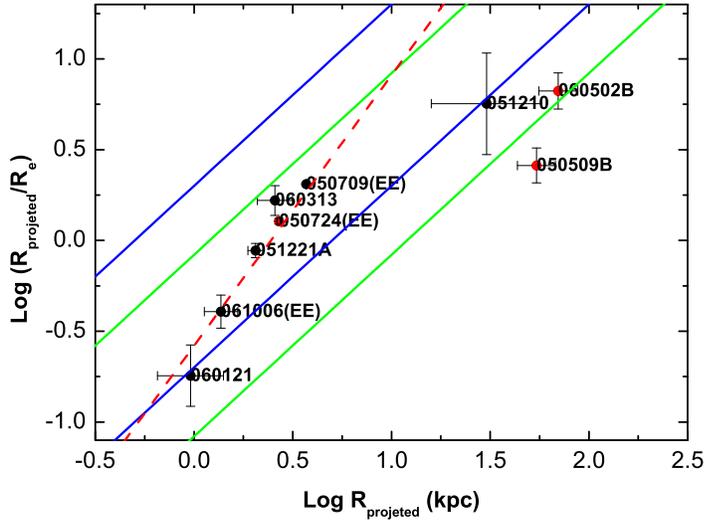}
\hfill \caption{Correlation of the Offset $R_{\rm{projected}}$ and
the Normalized Offset $R_{\rm{projected}}/R_{\rm{e}}$. The region
between solid lines are expected range for early-type host galaxies
in our calculation. The one between green lines are for late-type
galaxies. The red dashed line is the linear fitting to the bursts
excluding GRB050509B, GRB060502, and GRB051210. The bursts with EE
are noted behind the GRB name with ``EE''. The red dots are the
bursts with observed elliptical host galaxies. } \label{correlation}
\end{figure*}

We do the same analysis assuming that only one component (NN or NB)
can be BC. The results don't change so much, because the
distribution of NN and NB are very similar with each other.

We obtain the fraction of the SMC in this study, but what is the SMC
is a different discussion. It may be the collapsars, or young
neutron stars with strong magnetic fields (young magnetars), or
anything else. The SMC means the component that traces star-forming
regions. This discussion is out-scope of this study.

Our work is based on the star formation disk model (Bloom et al.
2002) and the PS model (Belczynski et al. 2006), whose uncertainties
will introduce uncertainties in our analysis, although it is claimed
that the most uncertain aspects are all parameterized to allow for
systematic error analysis (Belczynski et al. 2008). The smallness of
number of short GRBs whose locations in their host galaxies also
introduce uncertainties in our analysis. However, we would like to
note that our analysis is the first step to discuss the origin of
short GRBs from point of view of the Offsets/Normalized Offsets,
including the SMC in the analysis. We believe that this step is very
important to shed light on the origin(s) of short GRBs. We hope the
improvement of these models and increase of observed samples so that
we can discuss the origin of short GRBs with less uncertainties.

\section{Conclusions}

We have investigated the fractions of the SMC and BC as the origins
of short GRBs from the Offset/Normalized Offset analysis. We have
found that the fraction of the SMC is $0.37\pm0.13$ with error of
$1\sigma$ level for the Offset analysis, which suggests that the SMC
can be one of the origins of short GRBs, although most of short GRBs
are from BC. The fraction of the late-type host galaxies is
$0.82\pm0.05$ with error of $1\sigma$ level for the Offset analysis
of the Ellip-SB-Sp model, which is consistent with the observed
late- to early-type ratio of the host galaxies.

For the Normalized Offset analysis, the fractions of the SMC and
late-type host galaxies are $0.19\pm0.33$ and $0.30\pm0.21$,
respectively. We believe that the Offset analysis is more reliable
than the Normalized Offset analysis since the number of samples is
more limited for the Normalized Offset analysis. However, the
Normalized Offset analysis has a good potential to shed light on the
origins of short GRBs when number of samples becomes larger in the
future.

The fractions of the SMC are very similar for the bursts with EE and
those without EE, which suggests that their origins may be the same.
In the plane of the Offset and Normalized Offset, almost all of the
short GRBs in this study are in the expected ranges by our models.
There seems to be two groups in the plane, which may be related with
the type of host galaxies.

We thank for helpful discussion with the members of Kavli Institute
for Astronomy and Astrophysics and astronomy department of Peking
university. This research was supported by Grant-in-Aid for
Scientific Research on Priority Areas No. 19047004 by Ministry of
Education, Culture, Sports, Science and Technology (MEXT),
Grant-in-Aid for Scientific Research (S) No. 19104006 by Japan
Society for the Promotion of Science (JSPS), Grant-in-Aid for young
Scientists (B) No.19740139 by JSPS, Grant-in-Aid for Scientific
Research on Innovative Areas No. 21105509 by MEXT of Japan, and
Grant-in-Aid for the Global COE Program "The Next Generation of
Physics, Spun from Universality and Emergence" from MEXT of Japan.


\end{document}